\RequirePackage{fix-cm}
\documentclass[twocolumn,epjc3]{svjour3}  
\usepackage{graphicx}
\journalname{Eur. Phys. J. C}
\begin{document}
\title{Experimental search for the ``LSND anomaly'' with the ICARUS
  detector in the  CNGS neutrino beam}
\author{
M.~Antonello\thanksref{inst:1} \and 
B.~Baibussinov\thanksref{inst:2} \and 
P.~Benetti\thanksref{inst:3}  \and 
E.~Calligarich\thanksref{inst:3} \and 
N.~Canci\thanksref{inst:1} \and 
S.~Centro\thanksref{inst:2} \and  
A.~Cesana\thanksref{inst:4} \and  
K.~Cie\'slik\thanksref{inst:5} \and  
D.B.~Cline\thanksref{inst:6} \and 
A.G.~Cocco\thanksref{inst:7} \and 
A.~Dabrowska\thanksref{inst:5} \and 
D.~Dequal\thanksref{inst:2} \and 
A.~Dermenev\thanksref{inst:8} \and 
R.~Dolfini\thanksref{inst:3} \and 
C.~Farnese\thanksref{inst:2} \and 
A.~Fava\thanksref{inst:2} \and 
A.~Ferrari\thanksref{inst:9} \and 
G.~Fiorillo\thanksref{inst:7} \and 
D.~Gibin\thanksref{inst:2} \and 
S.~Gninenko\thanksref{inst:8} \and 
A.~Guglielmi\thanksref{inst:2} \and 
M.~Haranczyk\thanksref{inst:5} \and 
J.~Holeczek\thanksref{inst:10} \and 
A.~Ivashkin\thanksref{inst:10} \and 
M.~Kirsanov\thanksref{inst:8} \and 
J.~Kisiel\thanksref{inst:10} \and 
I.~Kochanek\thanksref{inst:10} \and 
J.~Lagoda\thanksref{inst:11} \and 
S.~Mania\thanksref{inst:10} \and 
A.~Menegolli\thanksref{inst:3} \and 
G.~Meng\thanksref{inst:2} \and 
C.~Montanari\thanksref{inst:3} \and 
S.~Otwinowski\thanksref{inst:6} \and 
A.~Piazzoli\thanksref{inst:3}  \and 
P.~Picchi\thanksref{inst:12} \and 
F.~Pietropaolo\thanksref{inst:2} \and 
P. Plonski\thanksref{inst:13} \and 
A.~Rappoldi\thanksref{inst:3} \and 
G.~L.~Raselli\thanksref{inst:3} \and 
M.~Rossella\thanksref{inst:3} \and 
C.~Rubbia\thanksref{inst:1,inst:9} \and 
P.~R.~Sala\thanksref{inst:4} \and 
E.~Scantamburlo\thanksref{inst:1} \and 
A.~Scaramelli\thanksref{inst:4} \and 
E.~Segreto\thanksref{inst:1} \and 
F.~Sergiampietri\thanksref{inst:14} \and 
D.~Stefan\thanksref{inst:1} \and 
J.~Stepaniak\thanksref{inst:11} \and 
R.~Sulej\thanksref{inst:11,inst:1} \and 
M.~Szarska\thanksref{inst:5} \and 
M.~Terrani\thanksref{inst:4} \and 
F.~Varanini\thanksref{inst:2} \and 
S.~Ventura\thanksref{inst:2} \and 
C.~Vignoli\thanksref{inst:1} \and 
H.G.~Wang\thanksref{inst:6} \and 
X.~Yang\thanksref{inst:6} \and 
A.~Zalewska\thanksref{inst:5} \and 
K.~Zaremba\thanksref{inst:13}
}                     
%
%
\institute{
INFN - Laboratori Nazionali del Gran Sasso, Assergi, Italy \label{inst:1}\and 
Universit\`a di Padova e INFN,  Padova, Italy\label{inst:2}\and
Universit\`a di Pavia e INFN, Pavia, Italy\label{inst:3}\and
Politecnico di Milano e INFN, Milano, Italy\label{inst:4}\and
H.Niewodnicza\'nski Institute of Nuclear Physics, Krak\'ow, Poland \label{inst:5} \and
Department of Physics, UCLA, Los Angeles, USA \label{inst:6} \and
Universit\`a Federico II di Napoli e INFN, Napoli, Italy \label{inst:7} \and
Institute for Nuclear Research of the Russian Academy of Sciences, Moscow, Russia \label{inst:8} \and
CERN, Geneva, Switzerland \label{inst:9} \and
A.So\l{}tan Institute for Nuclear Studies, Warszawa, Poland \label{inst:10} \and
Institute of Physics, University of Silesia, Katowice, Poland \label{inst:11} \and
INFN Laboratori Nazionali di Frascati, Frascati, Italy \label{inst:12} \and
Institute for Radioelectronics, Warsaw Univ. of Technology, Warsaw, Poland \label{inst:13} \and
Universit\`a di Pisa e INFN, Pisa, Italy \label{inst:14}
}
\date{Received: date / Revised version: date}
%
\maketitle
\begin{abstract}
We report an early result from the ICARUS experiment 
on the search for a  $\nu_\mu \rightarrow \nu_e$ signal 
due to the LSND anomaly. The search was performed  with the ICARUS 
T600 detector located at the Gran Sasso Laboratory, receiving  CNGS 
neutrinos from CERN at an average energy of about 20~GeV, 
after a flight path of $\sim$730~km.
The LSND  anomaly would manifest as an excess 
of $\nu_e$ events, characterized by a fast energy
oscillation averaging approximately to $\sin^2(1.27\Delta
  m^2_{new}L/E_\nu)$ $\approx 1/2$ with probability 
$ P_{\nu_\mu \rightarrow \nu_e} = 1/2
\sin^2(2\theta_{new})$. 
The present analysis  is based on 1091 neutrino
events, which are  about 50\% of the ICARUS data collected in 2010-2011.  
Two  clear $\nu_e$ events have been found, compared with the expectation of
$3.7\pm 0.6$ events from conventional sources.  Within the range of our
observations, this result is compatible with the absence of a LSND
anomaly.  At 90\% and 99\% confidence levels the limits of 3.4 and 7.3 events 
corresponding to oscillation probabilities $\left< P_{\nu_\mu
  \rightarrow \nu_e}\right> \le 5.4 \times 10^{-3}$ and $\left<
P_{\nu_\mu \rightarrow \nu_e}\right> \le 1.1 \times 10^{-2} $ are set 
respectively.
The result strongly limits the window of open options for the LSND anomaly  
to a narrow region around $\left( \Delta m^2 , \sin^2{(2 \theta )} \right)_{new} = (0.5\ 
\mathrm{eV}^2, 0.005)$, where there is an overall agreement (90\% CL)
between the present ICARUS limit, the published limits of KARMEN and
the published positive signals of LSND and MiniBooNE Collaborations.
\PACS{
      Neutrino Ocillations   \and
      Sterile neutrinos
     } 
\end{abstract}
%
%
\section{Introduction}
\label{sec:intro}
The possible presence of neutrino oscillations into sterile states has been
proposed by B. Pontecorvo~\cite{Pontecorvo}. An experimental search
for an anomalous $\bar \nu_e$ production at short distances has been
performed by the LSND experiment~\cite{LSND} at the Los Alamos 800 MeV
proton accelerator, which reported an anomalous excess of $\bar \nu_e$  from
$\bar \nu_\mu$ originated by muons from pions at rest with $\left<
E_\nu \right> \approx$~30~MeV and $L\approx$~30~m.
It is well known that  anti-neutrino oscillations at such a small distance
from the source should imply the presence of additional
mass-squared differences, largely in excess of the three neutrino mixing 
standard model values. The LSND signal  $\left< P_{\bar \nu_\mu \rightarrow \bar 
\nu_e}\right> = (2.64 \pm 0.67 \pm 0.45) \times 10^{-3}$ corresponds
to a rate of $(87.9 \pm 22.4 \pm 6.0)$ events, namely a 3.8~$\sigma$
effect at $L/E_\nu \sim$ $0.5-1.0$~m/MeV.

A recent result from MiniBooNe~\cite{MiniBoone}, performed with
neutrinos from the 8 GeV FNAL-Booster in a similar $L/E_\nu$ range has
confirmed in both the neutrino and antineutrino channels a combined
$3.8\ \sigma$ LSND-like oscillation signal. With the formula
\begin{equation}
\left< P_{\nu_\mu \rightarrow \nu_e}\right> = \sin^2{ \left(
2 \theta_{new}\right)} \sin^2{\left( 
\frac{1.27\Delta m^2_{new} (\mathrm{eV}^2) L(\mathrm{m})}{E_\nu (\mathrm{MeV})} \right )}
\end{equation}
 these results correspond to a new signal somewhere within
a wide interval $\Delta m^2_{new}\approx$~0.01 to 1.0~eV$^2$ and a
corresponding associated value of $\sin^2{\left( 2  \theta_{new}\right)}$.

In addition,  an apparent $\nu_e$ or $\bar \nu_e$ 
disappearance anomaly has been recently detected from (a) nearby nuclear
reactors~\cite{Reactors} and (b) from Mega-Curie k-capture calibration
sources~\cite{MegaCurie,Gallex}, originally developed for the Gallium
experiments to detect solar $\nu_e$. Also these effects seem to occur  
for a $\Delta m^2_{new}$ value much higher than  the experimentally measured 
ones for the three neutrino oscillation scenario, in the order of magnitude 
of the LSND anomaly. These anomalies
may indeed represent an
unified approach, in which one or more $\Delta m^2_{new}$ may have a
common origin, with the values of $\sin^2{\left( 2\theta_{new}\right)}$ for 
different channels reflecting the so far unknown structure of the 
${\bf U}_{(j,k)}$ matrix, with $j,k$~=~number of ordinary and sterile neutrinos.
  
 With the help of a novel development of a large mass ``Gargamelle 
 class'' LAr-TPC imaging detector, the ICARUS 
 experiment~\cite{ICARUS-INAUGURAL,ICARUS-BIBBIA} at the Gran Sasso 
 underground laboratory (LNGS) is hereby visually searching 
 for the signature of such a signal due to a LSND-like anomaly in the 
 CERN to Gran Sasso neutrino beam (CNGS).

\section{The experimental setup}
\label{sec:exper}

The CNGS facility~\cite{CNGS} provides a neutrino beam composed mainly of
muon neutrinos peaked in the range $ 10 \le E_\nu \le
30$~GeV. The CERN-SPS 400 GeV proton beam with about $2 \times
10^{13}$ protons on target (pot) per spill is sent to a segmented carbon target followed by a
magnetic horn and a reflector, focusing charged secondary mesons into
a 1~km long decay tunnel.  Produced neutrinos are pointing down with  a
52~mrad slope toward the Gran Sasso  laboratory (LNGS) located at a distance of 730 km.

According to detailed Monte Carlo (MC) calculations of the  neutrino
beam~\cite{CNGS-sim}, about 2850 charged current (CC) events/kt/year are expected at LNGS for a
nominal proton beam intensity of $4.5 \times 10^{19}$ pot/year with a
spectral contamination from anti-neutrino of about 2\% and an
electron component of slightly less than 1\%. The neutrino
flux and spectra expectations are obtained with a complete simulation of
all the beam line elements based on the FLUKA Monte Carlo
code~\cite{FLUKA1,FLUKA2}, and are available to all experiments on the
CNGS beam~\cite{nufluxweb}. The hadron interaction models in FLUKA have been
benchmarked on several sets of experimental data, among which the data
from the old NA20, NA56 and the present NA49 hadron production
experiments~\cite{Atherton,SPY,Collazuol,NA49}. 
Conservatively a 10 \%
systematics, introduced by the hadron production model in the computed
fluxes, can be assessed when averaging over the angular
acceptance of $\approx$~30~mrad of the beam optics. This level of
agreement is demonstrated, for example, by the comparison with
p$_t$-integrated pion production data of NA49  as 
reported in Figure~7 in~\cite{FLUKA2}.

This conclusion is corroborated by the absolute comparison of the
horizontal and vertical distributions of the signals of the CNGS muon
pit detectors with the full beam line simulation, have shown an agreement
within few percents in the first pit and ranging from few percents to
10\% in the second one~\cite{CNGS1}.
 
According to the full neutrino beam calculation, 75\% of muon neutrinos 
are coming from decays of pion produced at the target, the rest is due to 
kaons, (6\%) and tertiary decays (19\%).  
Electron neutrinos are originated by pions through the subsequent muon decay (37\%) as well as
by kaons (43\%), the remaining 20\% is due to tertiary decays.  
Hence, due to correlations between the $\nu_\mu$ and $\nu_e$
common origins, significant cancellations occur in the systematics of the
$\nu_e / \nu_\mu $ ratio. As a result, the integral error on the $\nu_e/
\nu_\mu $ ratio is estimated to be better than 7~\%.

The ICARUS experiment is operated at  $L/E_\nu\approx 36.5$~m/MeV, 
a value much larger than the
one of the  experiments where anomalies appeared.
 In first approximation, a hypothetical $\nu_\mu
\rightarrow \nu_e $ LSND anomaly will  produce very fast
oscillations as a function of the neutrino energy $E_\nu$, averaging
to $\sin^2{\left(1.27\Delta m^2_{new}L/E_\nu \right )} \approx 1/2$ and
$\left< P_{\nu_\mu \rightarrow  \nu_e}\right>=1/2\sin^2{\left(2\theta_{new}\right)}.$  
This signal
will have to be compared with the small, but significant, backgrounds
due to other and more conventional neutrino sources.

It is  well known and widely described in~\cite{ICARUS-BIBBIA}, 
that the TPC developed
by the ICARUS group provides, in a massive liquid Argon (LAr) volume, a
completely uniform imaging with accuracy, density and interaction
lengths comparable to the ones  of, for instance, a  heavy Freon bubble
chamber. This innovative detection technique  allows observing  
the actual ``image'' of each charged
track with a resolution of few mm$^3$, thus extending in a liquid the
method originally proposed by Charpak et al.~\cite{Charpak} in a
gas. 

The ICARUS-T600 detector, smoothly operated 
over the last three years 
in the underground Hall B of the LNGS laboratory,  has a mass in excess of 600 ton of ultra high
purity LAr, out of which 476 are instrumented and 447 are defined as
fiducial volume for the selection of neutrino events.
A detailed description of the detector design,
construction and test can be found in dedicated
articles~\cite{ICARUS-INAUGURAL,ICARUS-BIBBIA}. It allows identification and measurement of
the ionisation image of all tracks produced within the fiducial
volume to which a 500~V/cm uniform electric field is applied (for maximum drift path of 1.5~m).  
Sensing and recording of the signals induced by the drifted
electrons (drift velocity $\approx$ 1.6 mm/$\mu$s) is provided by a set 
of three parallel planes of wires, 3~mm apart, 3~mm pitch, facing the drift volume.  
Wires  on each plane are oriented at a different angle
 (0$^\circ$, +60$^\circ$, -60$^\circ$) with respect to the
horizontal direction. By appropriate voltage biasing, the first
two planes (Induction-1 and Induction-2) provide signals in
non-destructive way, whereas the ionisation charge is finally
collected by the last one (Collection). 
This provides three projective views of the same event simultaneously, 
allowing both space point reconstruction and precise calorimetric 
measurement of the collected charges.
 
In order to ensure in LAr the visibility of tracks drifting over 
several meters, an equivalent Oxygen electro-negative content smaller  
than  a few tens of ppt (parts per trillion) is required.  During
the present experiment, the free electron
lifetime has been maintained most of the time in excess of 5 ms
corresponding to a maximum 18\% signal correction for the longest 1.5~m drift path 
 of the LAr-TPC~\cite{ICARUS-INAUGURAL}.

Electronics is designed to allow continuous read-out, digitization and
independent waveform recording of signals from each of the wires of
the TPC. A 10-bit ADC digitization at 400 ns sampling provides a
dynamic range of up to about 100 minimum ionising particles.  The
average electronic noise is typically of about 1500 electrons r.m.s.,
to be compared with $\sim$15000 free electrons signal
recorded for a 3 mm minimum ionising particle (S/N$\sim$10).

A total of 74 photomultipliers (PMT) of 8'' diameter sensitive to the
128 nm LAr UV-light, located behind the transparent wire planes, are
used to detect the prompt scintillation light 
produced in LAr simultaneously with ionisation.
They are used to trigger the presence of the neutrino signal  within 
a CNGS related 60 $\mu$s gate and define the precise location of the 
event along  the drift direction~\cite{ICARUS-INAUGURAL},~\cite{nutof1}. 
 A PMT threshold,  set at 100 photoelectrons, allows full detection 
 efficiency for events with energy deposition (E$_{dep}$) as low as 
 few hundreds MeV.  Indeed, a trigger efficiency exceeding 99$\%$ 
 for E$_{dep} > $~500~MeV has been measured on a large 
 event sample ($1.1 \times 10^6$ spills, $1.7 \times 10^{19}$ p.o.t.) 
 collected triggering only on the CNGS extraction signal. 
The trigger
efficiency at very low energies depends on the topology and
localisation of the event. Monte Carlo simulations indicate a 100\%
efficiency for CNGS charged current (CC) events and a $>$90\% efficiency for
neutral current (NC) events. 

A CNGS trigger rate of about  1 mHz was obtained including neutrino 
interactions inside the detector and muons from neutrino interactions 
in the upstream  rock. 

\section{Data selection}
\label{sec:selec}

\begin{figure}[tbh]
\centering
\resizebox{0.4\textwidth}{!}{\includegraphics{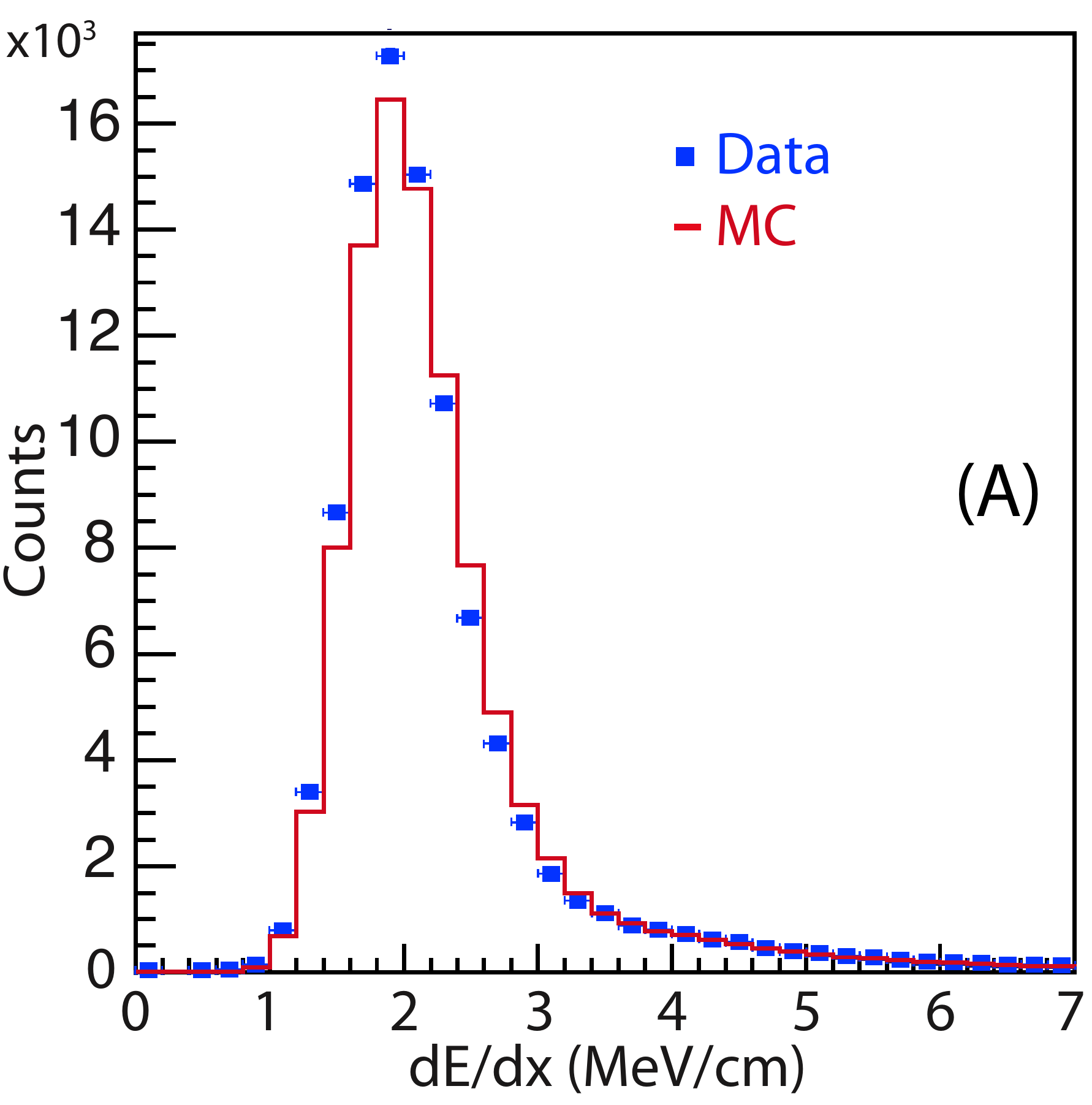}}
\resizebox{0.4\textwidth}{!}{\includegraphics{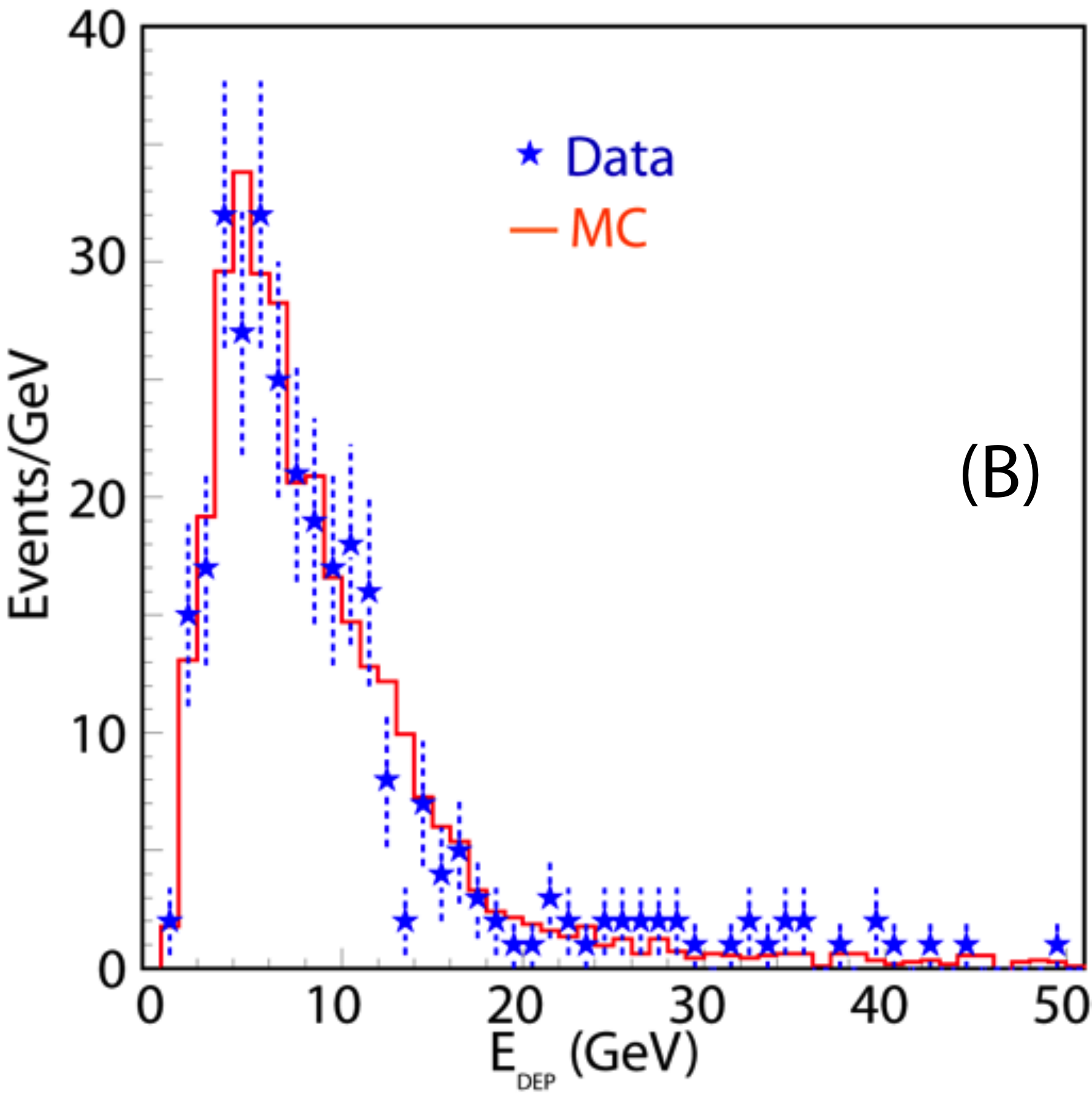}}
\caption{(A) Energy deposition density distribution for muons in 
CNGS CC interactions, compared with Monte Carlo, normalised to the same number of entries. 
Each entry corresponds to one wire hit.
(B) Experimental raw energy distribution $E_{dep}$ for muon 
neutrinos and antineutrinos CC interaction in the ICARUS T600 
detector (blue symbols) compared with the Monte Carlo expectations 
(red solid histogram), normalised to the same number of entries.}
\label{fig:comp}
\end{figure}

\begin{figure}[tbh]
\centering
\resizebox{0.45\textwidth}{!}{\includegraphics{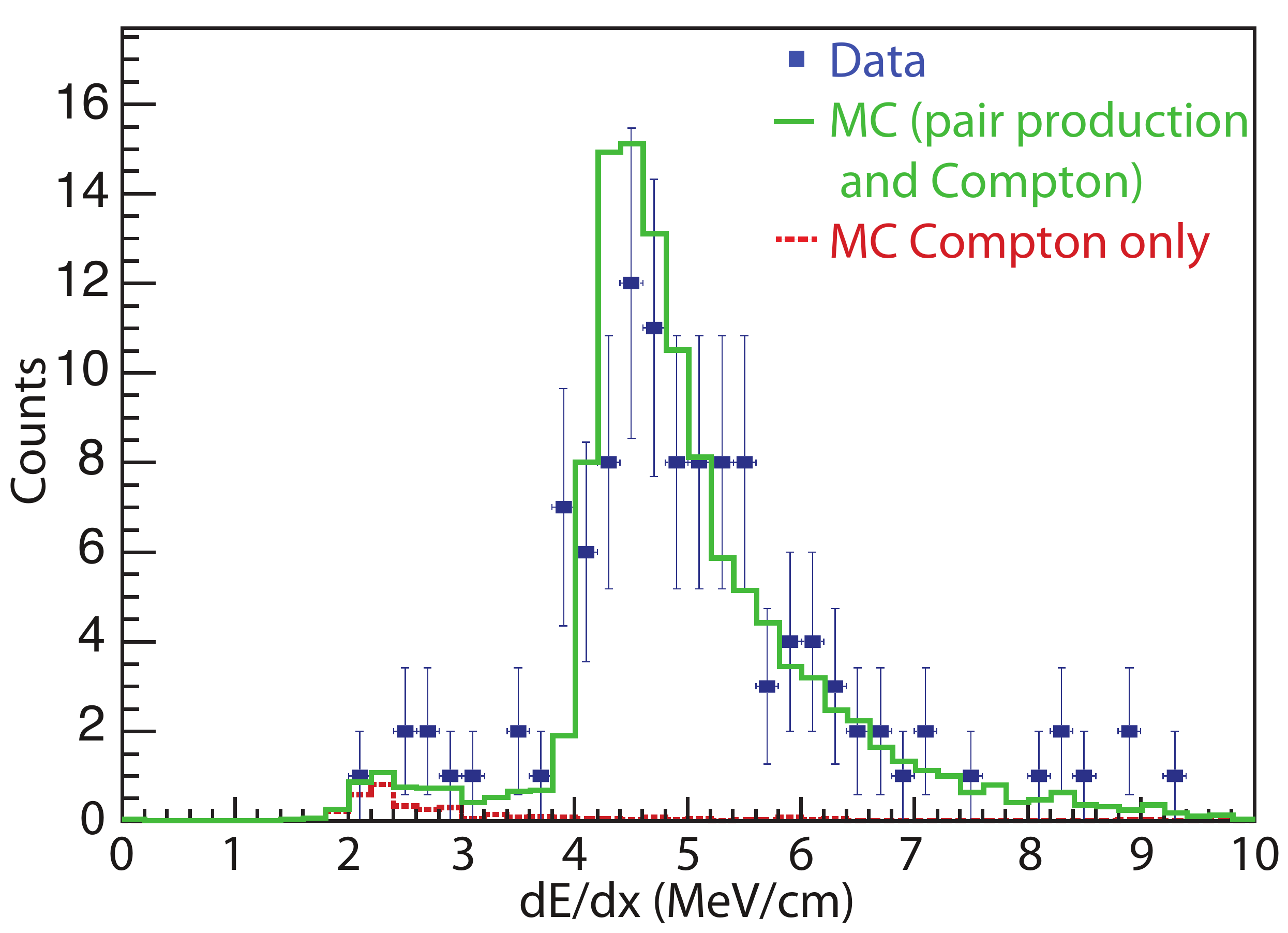}}
\caption{Average ionisation in the first 8 wire hits 
for sub-GeV photons in the T600 data (full squares), compared to Monte Carlo expectations 
(solid line) normalised to the same number of events. 
In MC case, the Compton contribution is shown also separately (dotted line).}
\label{fig:showers} 
\end{figure}

\begin{figure}[tbh]
\centering
\resizebox{0.45\textwidth}{!}{\includegraphics{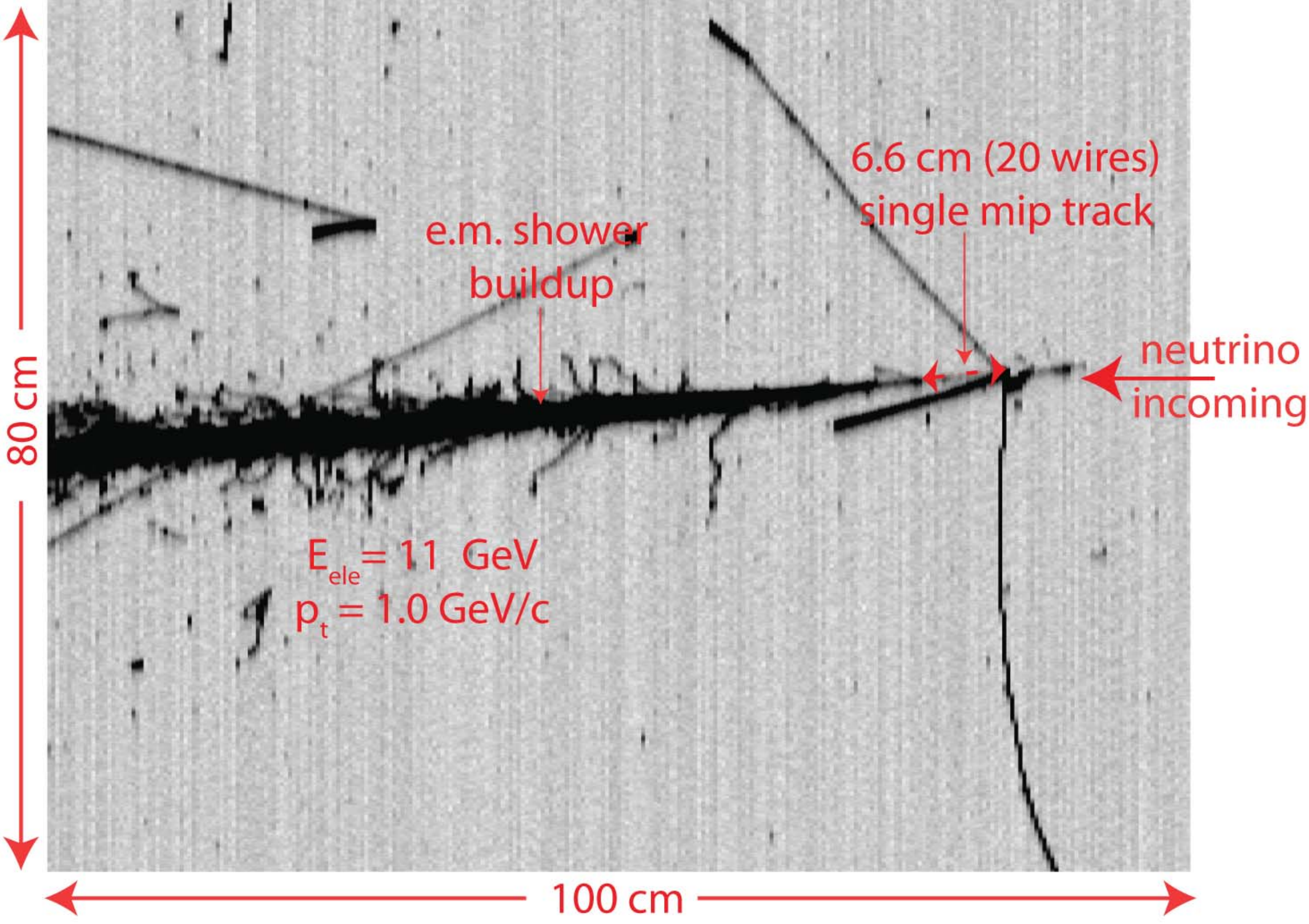}}
\caption{Typical Monte Carlo generated $\nu_\mu
\rightarrow \nu_e$ event from the ICARUS full simulation 
program~\protect\cite{FLUKA1,FLUKA2,FLUKA-nu} with $E_e = 11$~GeV and $p_T = 1.0$~GeV/c. 
The close similarity of the MC simulation with actual ICARUS events 
(see Figure~\protect\ref{fig:eventi} (A) and (B)) is apparent.}
\label{fig:MCev} 
\end{figure}

Empty events inside the recorded CNGS sample are rejected through a 
dedicated automatic filter based on charge deposition, whose efficiency 
close to 100\% has been checked on a sample of few thousands visually  
scanned events. A few neutrino interactions/day with vertex in the fiducial 
volume are recorded, as expected.
 
The identification of the primary vertex and of 2D objects, like
tracks and showers,  is
performed visually. The  obtained clusters and reference points are
fed to the three dimensional reconstruction algorithm
described in detail in~\cite{3Dpaper}. The collected charge   is
calculated for each ``hit'' (a point in the wire-drift projection) in
the Collection view  after automatic hit finding and hit
fitting~\cite{ICARUS-BIBBIA},~\cite{3Dpaper}. Each hit is corrected for the signal 
attenuation along the drift, according to the purity value as continuously monitored with cosmic muons.
Stopping tracks are processed
for particle identification through specific
ionisation~\cite{3Dpaper}. 
The total deposited energy is obtained by calibrated 
sum of hit  charges   in  the region spanned by the event,  with an 
average correction factor  for  signal quenching in LAr.  
   Muon neutrino charged current
events are identified with the requirement of a track exiting the primary
vertex and travelling at least 250~cm in the detector.    

In order to reproduce  the signals from the  actual events,
a sophisticated simulation package dedicated to the ICARUS T600 detector
has been developed. Neutrino events are generated according to the
expected spectra with uniform vertex position within the T600
sensitive volume. The adopted neutrino event generator~\cite{FLUKA-nu}
 includes quasi-elastic, resonant and deep inelastic
processes and is embedded in the nuclear reaction model of FLUKA.
Therefore it accounts for the effects of Fermi motion,  Pauli
principle, and  other initial and final state effects such as, for
instance, reinteractions of the reaction products inside the target
Argon nucleus~\cite{50l}.  All reaction products are transported in the T600
volume, with detailed simulation of energy losses by ionisation, delta
ray production, electromagnetic and hadronic interactions.  Ionisation
charge along the track is subject to the experimentally observed
recombination effects~\cite{LAR1}.  Energy depositions are registered
in grid structures that reproduce the actual wire orientation and
spacing, with a fine granularity (0.2~mm) in the drift
direction. The resulting charge is convoluted with the readout channel
response (including wire signal induction and electronics response),
including noise parameters extracted from the real data.  Such a procedure results
in a remarkably close similarity between real and simulated events.

The good agreement 
between the observed and predicted wire
signals is shown in Figure~\ref{fig:comp}A for
CNGS muon tracks recorded in  CC
events. Those tracks are distributed over all the detector volume, 
and have been  recorded during  several months of operation, 
thus this plot includes all possible effects due to spatial or temporal non-uniformity.
The agreement on the average value is at the level of 2.5\%. Similar
comparisons have been performed on single tracks from long stopping
muons, whose energy can be measured. The distribution of dE/dx for
each track has been fitted with the convolution of a Landau function
with a gaussian. The fitted value of the  most probable dE/dx agrees
at 2\% level with Monte Carlo expectations, and the fitted  gaussian $\sigma$ 
is about 10\%, reflecting the expected hit charge signal/noise ratio~$\sim$~10.
Similar agreement is obtained for protons and
pions~\cite{3Dpaper}.  Figure~\ref{fig:comp}B shows the experimental raw energy
distribution $E_{dep}$ for the observed $\nu_\mu + \bar\nu_\mu$ CC interactions
compared with the MC expectations~\cite{LAR2}. The average value of the energy 
deposited in the detector is reproduced within  2.5\% and its rms  within  10\%.

\begin{figure*}[tbh]
\centering
\resizebox{0.45\textwidth}{!}{\includegraphics{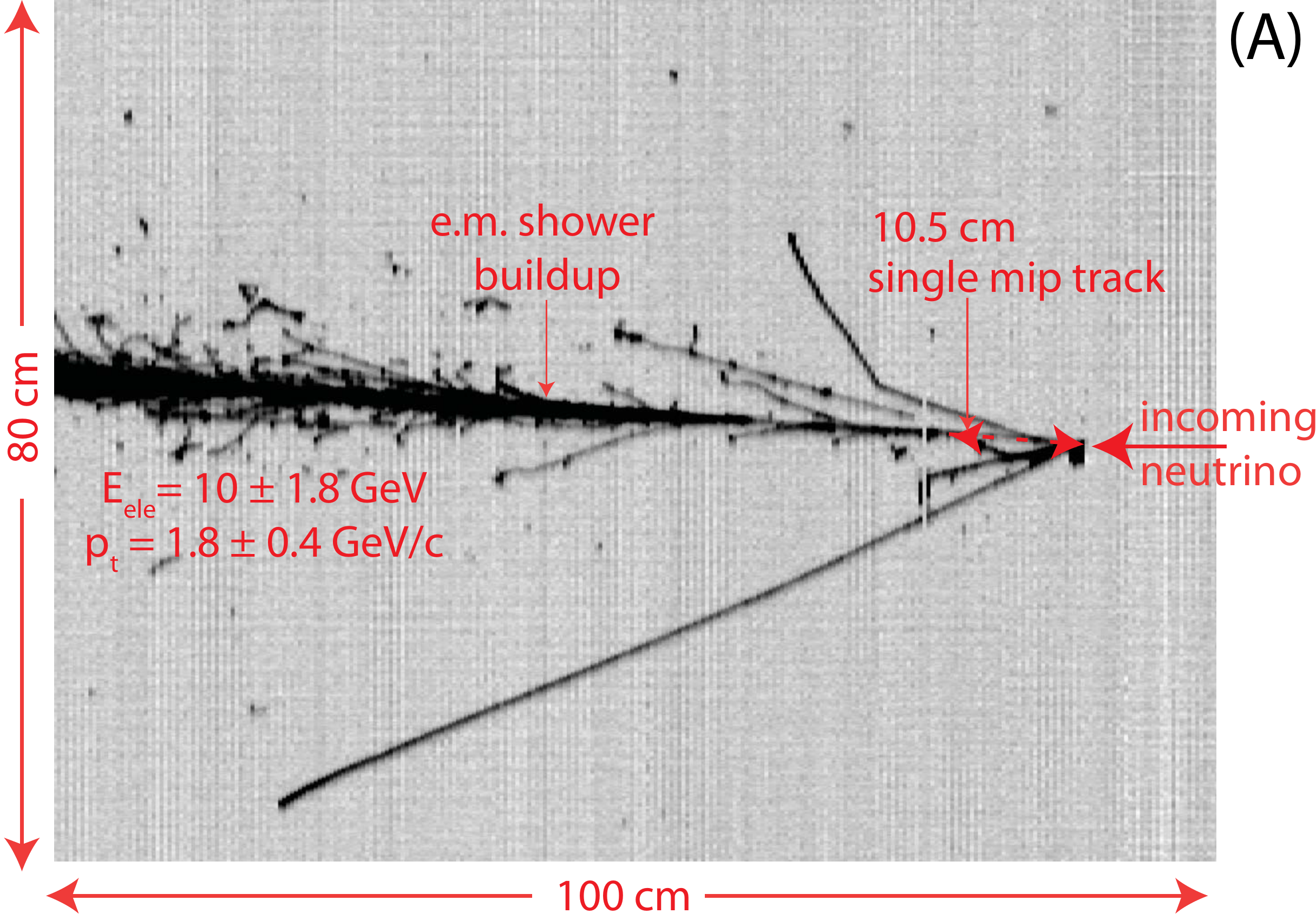}}
\resizebox{0.45\textwidth}{!}{\includegraphics{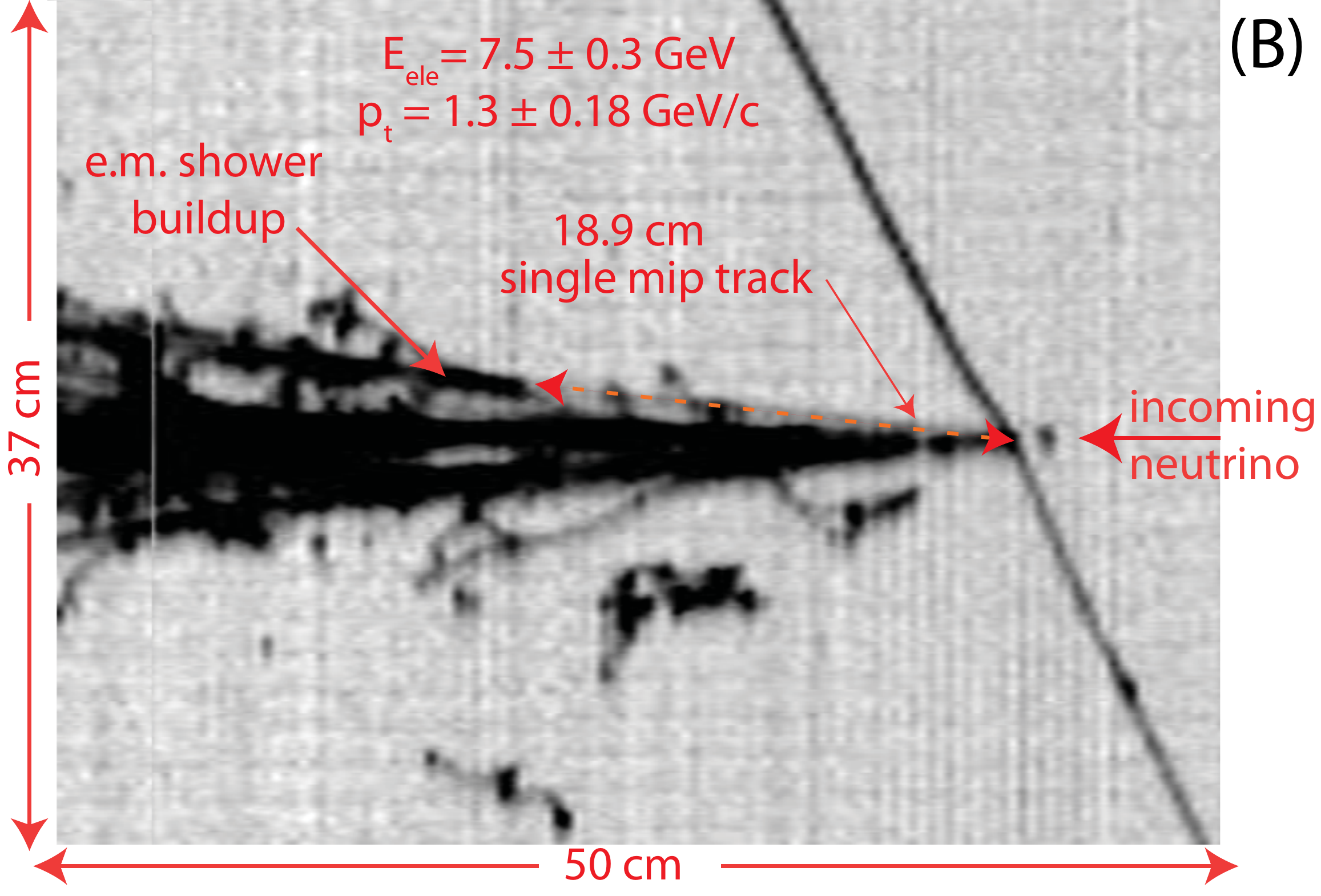}}
\resizebox{0.70\textwidth}{!}{\includegraphics{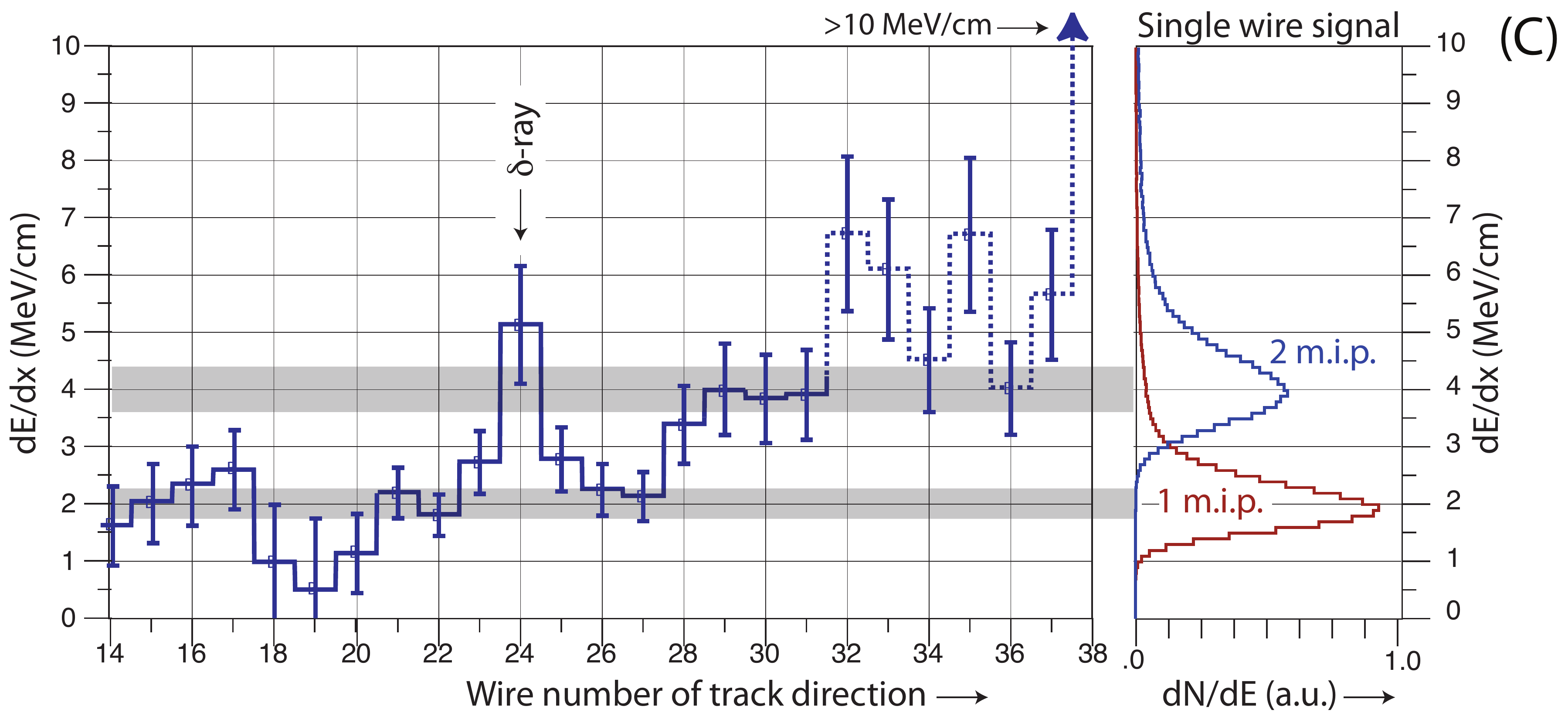}}
\caption{Experimental picture of the two observed events (A) and (B) with a
clearly identified electron signature out of the total sample of 1091
neutrino interactions. Event in (A) has a total energy of $11.5 \pm
1.8$~GeV, and a transverse electron momentum of $1.8 \pm 0.4$~GeV/c.
Event in (B) has a visible energy of $\sim$17 GeV and a transverse
momentum of $1.3 \pm 0.18$~GeV/c.  In both events the single electron
shower in the transverse plane is clearly opposite to the remaining of
the event.  (C): display of the actual dE/dx along individual wires of
the electron shower shown in Figure~\ref{fig:eventi}A, in the region
($\ge$ 4.5~cm from primary vertex) where the track is well separated from
other tracks and heavily ionising nuclear prongs. As a reference, the
expected dE/dx distribution for single and double minimum ionising
tracks (see Figure~\ref{fig:comp}A), are also displayed.  The dE/dx
evolution from single ionising electron to shower is also shown.}
\label{fig:eventi} 
\end{figure*}

The search for $\nu_\mu \rightarrow \nu_e$ events due to a LSND
anomaly has been performed as follows. The ICARUS experimental sample
has been based on 168 neutrino events 
collected in 2010 ($5.8 \times 10^{18}$ pot) and 923 events collected in 2011 ($2.7 \times
10^{19}$ pot out of the $4.4 \times 10^{19}$ collected in 2011),
leading to a total of 1091 observed
neutrino events,  in good agreement, 
within 6\%, with the Monte Carlo expectation.
To this initial sample, a minimal fiducial volume cut has been applied  to collect as
much statistics as possible: the interaction vertex is required to be
at a distance of at least 5 cm  from each side of the
active volume and at least 50 cm  from its  downstream wall.  These cuts allow
for the identification of electron showers, but are neither stringent
enough for the reconstruction of neutrino energies, nor for the
identification of $\nu_\mu$CC vs NC events. 
Furthermore, only events with a deposited energy smaller
than 30~GeV have been included in the analysis, in order to optimize the signal over background
ratio. Indeed, the oscillated events are expected to have energies in
the 10-30 GeV range, like the bulk of the muon neutrino spectrum,
while the 
beam $\nu_e$ contamination extends to higher energies. 

 The estimation of the
fraction of background and oscillated events falling in the required
energy cuts has been performed on large samples (order of 10000 for
each neutrino specie) of simulated events,
where the spectrum of the oscillated events has been assumed to be
equal to the $\nu_\mu$~CC spectrum (mass effects on the cross
sections are assumed to be negligible in this energy range).  
Since the agreement of the simulations with the deposited energy
spectrum is very good and the energy cut concerns only about 15\% of
the events, the cut on visible energy introduces a negligible
systematic error on the signal expectation. The same is true for all
background sources, except the 
$\nu_e$ beam component whose
energy spectrum extends to higher energies. In this case, any
uncertainty in the deposited energy spectrum is reflected in an equal
uncertainty on the effect of the energy cut. 
On the basis of the comparisons
 shown in in Figure~\ref{fig:comp} and described previously, 
 we assumed a conservative 10\%
systematics on the effect of the energy cut on the 
beam $\nu_e$ background, to be added to the one
on the prediction of the $\nu_e/\nu_\mu$ ratio.

All Monte Carlo predictions have been normalized to the experimental total 
number of observed CNGS neutrino events before any cut.

The radiation length of LAr is 14 cm ($\approx 45$ readout wires),
corresponding to a $\gamma$-conversion length of 18 cm. The ionisation
information of the early part -- before the showering of the
e.m. track has occurred -- is examined wire by wire in order to tag
the presence of an initial electron emitted in the neutrino interaction, as
a powerful eliminator of $\gamma$-converting pairs, which are generally
separated from the vertex and generate double minimum ionising tracks.
The rejection factor based on ionisation increases dramatically with
increasing photon energies, while the electron identification
efficiency is almost constant.  Indeed, the possible photon
misidentification is essentially due to photons undergoing Compton
scattering, whose cross section becomes negligible with respect to the
pair production above a few hundreds MeV. Monte Carlo studies indicate
a residual contamination of about 0.18\% for the energy spectrum of photons  
from  pion decays in CNGS events, rising to a few \% in the sub-GeV energy region. 
The loss in efficiency for electron showers is only 10\%.  First results from an 
ongoing study on low energy showers from isolated secondary $\pi^0$'s in the 
T600 CNGS data confirm the MC expectation (see Figure~\ref{fig:showers}). 
The plot shows a good agreement between 
data and simulations, including the low ionisation tail due to Compton interactions.

In the present analysis, the ``electron signature'' has been defined by the following
requirements: 
\begin{itemize}
\item[(a)] vertex of the event inside the fiducial volume; 
\item[(b)] visible event energy smaller than 30 GeV, in order to reduce the
beam $\nu_e$  background;
\item[(c)] the presence of a charged track starting directly from the vertex, fully
consistent over at least 8 wire hits with a minimum ionising
relativistic particle, i.e. the average 
dE/dx must be  lower than 3.1 MeV/cm after removal of visible delta rays (see Figure~\ref{fig:comp}A),   and
subsequently building up into a shower; 
\item[(d)] visible spatial separation
from other ionising tracks within  150~mrad in the immediate vicinity
of the vertex in at least one of the two transverse views ($\pm
60^\circ$), except for short proton like recoils due to nuclear
interactions.
\end{itemize}

In order to determine the electron signature selection efficiency $\eta$, 
$\nu_e$ events have been generated
with MC according to the $\nu_\mu$ CC spectrum.
 A simulated event is shown in Figure~\ref{fig:MCev}. Out of
an initial sample of 171 $\nu_\mu \rightarrow \nu_e$ MC reconstructed
events, 146 events have a visible energy smaller than  30 GeV, 122 of which 
satisfy the fiducial volume cuts (a). These events have been visually
and independently scanned by three different people in different
locations.  An excellent agreement has been found with differences in
less than 3\% of the sample. 
As a result,
the average number of positively identified electron-like neutrino
events is 90,
corresponding to a selection efficiency $\eta=0.74\pm0.05$.  
In a good approximation $\eta$ is independent of the details of the energy
spectrum. The systematic error on $\eta${} induced by the dE/dx cut is bound 
to be smaller than 1\% from the already discussed agreement to better than
2.5\% between the measured and the predicted scale of the 
dE/dx for muons in $\nu_\mu CC${} (see Figure~\ref{fig:comp}A).

A similar scan of 800 MC neutral current events has shown no presence of
apparent $\nu_\mu \rightarrow \nu_e$ events, consistent for our sample
with an estimated upper limit of 0.3 events (including possibly misidentified 
$\nu_\mu$ CC events).  Moreover, an independent
estimation of the background rejection efficiency has been performed
on a much larger MC sample with a fast simulation and reconstruction
algorithm. All CNGS beam original and oscillated neutrino flavors have
been taken into account. Automatic cuts mimicking the data cuts have
been applied to the simulated events. After the fiducial and deposited
energy cuts (C1 in Table~\ref{table:MCcutsT}) , background neutral current 
and charged current events
have been retained as ``electron'' candidates if no muon-like track
could be indentified, and at least one energetic photon (at least 100
MeV) pointing to the primary vertex was present (C2). The requirements for
the shower isolation and for a conversion distance smaller than 1~cm
were then applied (C3). Finally, the discrimination based on the specific
ionisation was applied as an average factor (C4).  The effect of
the various cuts is summarised in Table~\ref{table:MCcutsT}.
With this method, that is not fully equivalent to the visual scan, 
the estimated background from
misidentified NC and $\nu_\mu$ CC events amounts to 0.09 events, and the  
the simulated efficiency on $\nu_e$ CC events (after the fiducial and energy cuts) 
is found to be  74\% in agreement with the scanning method. 
The contribution from a 2.5\% uncertainty on the
dE/dx scale would modify this background estimate by less than 10\%.

\begin{table}[tbh]
\centering
\caption{Fraction of Monte Carlo events surviving the automatic selection cuts, defined as follows. 
  C1: $E_{dep} < 30$~GeV; C2: no identified muon, at least
  one shower;  C3: one shower with initial point (conversion point
  in case of a photon) at a distance smaller than 1 cm from the
  neutrino interaction vertex, separated from other tracks; C4: single
  ionisation in the first 8 samples.  All event categories are reduced to 0.93
  after the  cut on fiducial volume. The signal selection efficiency (after the fiducial and energy cuts)
  results to be 0.6/0.81=0.74, in agreement with the visual scanning method.}
\label{table:MCcutsT}
\begin{tabular}{|c|c|c|c|c|c|c|}
\hline
&&&&&&\\
Sel.  & $\nu_e$ CC & $\nu_e$ CC & $\nu_\tau $ CC &    
                   NC        &  $\nu_\mu$ CC            & $\nu_e$ CC\\
cut &  beam & $\theta_{13}$ &  &&& signal  \\
\hline
&&&&&&\\
 C1    &    0.47   &  0.92   &   0.93   & 0.89        & 0.89 &  0.81 \\    
 C2    &    0.47   &  0.92   &   0.17   & 0.66        & 0.19 &  0.81 \\
 C3    &    0.33   &  0.79   &   0.14   & 0.10        & 0.03 &  0.66 \\
C4    &    0.30   &  0.71   &   0.13   &0.0002 &0.00005 & 0.60\\
&&&&&&\\
\hline
\end{tabular}
\end{table}

The expected number of $\nu_e$ events due to conventional sources in
the energy range and fiducial volumes defined in (a) and (b) are as
follows: 
\begin {itemize}
\item  $3.0 \pm 0.4$ events due to the  estimated 
$\nu_e$ beam  contamination; 
\item  $1.3 \pm 0.3$ $\nu_e$
events due to the presence of $\theta_{13}$ oscillations from
$sin^2(\theta_{13}) = 0.0242 \pm 0.0026$~\cite{FOGLI:t13};
\item   $0.7 \pm 0.05$
$\nu_\tau$ with $\tau \rightarrow e$ from the 
three neutrino mixing standard model 
predictions~\cite{PDG}, 
\end{itemize}
giving a total of $5.0 \pm 0.6$ expected
events, where the uncertainty on the NC and CC contaminations has been included.
 The expected visible background  is then $3.7 \pm 0.6$
(syst. error only) events
after the selection efficiency $\eta=0.74\pm0.05$ reduction has been applied.
Given the smallness of the number of electron like signal expected in absence
of LSND anomaly, the estimated systematic uncertainty on the predicted number 
is clearly negligible w.r.t. its statistical fluctuation. 

In the recorded experimental sample, two events in which a 
$\nu_e$ signature have been identified, to be compared with the above
expectation of 3.7 events for conventional sources.  The event in
Figure~\ref{fig:eventi}A has a total energy of $11.5 \pm 2.0$ GeV and
an electron of $10 \pm 1.8$ GeV taking into account a partially
missing component of the e.m. shower.  The event in
Figure~\ref{fig:eventi}B has 17 GeV of visible energy and an electron
of $7.5 \pm 0.3$ GeV.  In both events the single electron shower in
the transverse plane is opposite to the remaining of the
event, with the electron transverse momentum of $1.8 \pm 0.4$ GeV/c
and $1.3 \pm 0.18$ GeV/c respectively.

Figure~\ref{fig:eventi}C displays the actual dE/dx along individual
wires of the electron shower shown in Figure~\ref{fig:eventi}A, in the
region ($\ge$~4.5~cm from primary vertex), where the track is well separated
from other tracks and heavily ionising nuclear prongs. As a reference,
the expected dE/dx distribution for single and double minimum ionising
tracks (see Figure~\ref{fig:comp}A), are also displayed.  The dE/dx
evolution from single ionising electron to shower is also shown.

\section{Results and discussion}
\label{sec:resul}

\begin{figure}[tbh]
\centering
\includegraphics[width=0.45\textwidth]{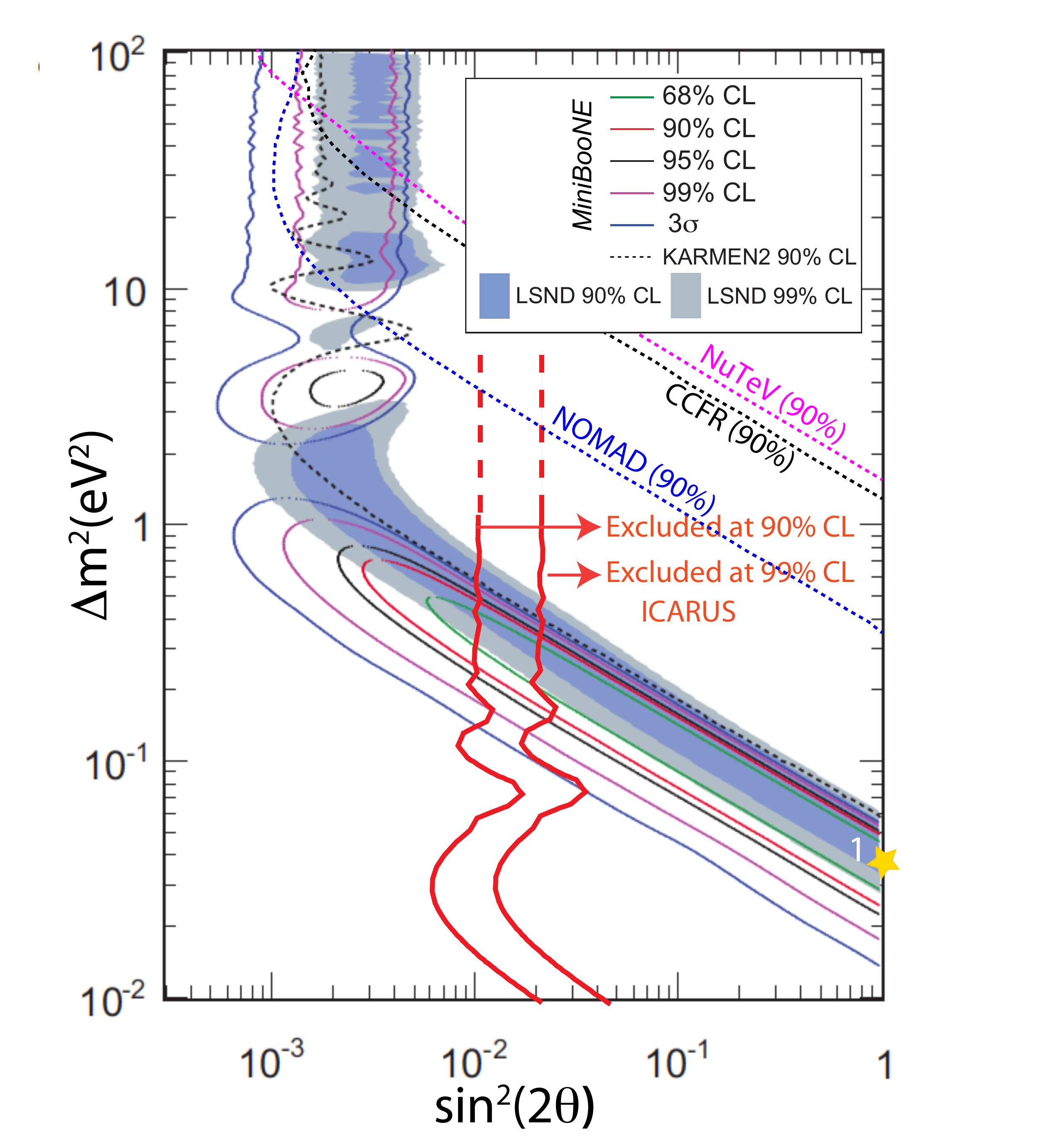}
\caption{Two-dimensional plot of $\Delta m^2$ vs $\sin^2{(2 \theta_{new} )}$ 
for the main published experiments sensitive to the $\nu_\mu
\rightarrow \nu_e$ anomaly~\protect\cite{LSND,MiniBoone,KARMEN,NOMAD,CCFR,NuTeV}
and the present ICARUS result. The ICARUS limits to the oscillation probability 
are  $\left< P_{\nu_\mu \rightarrow \nu_e}\right> \le 5.4 \times 10^{-3}$ and $\left<
P_{\nu_\mu \rightarrow \nu_e}\right> \le 1.1 \times 10^{-2} $~,
corresponding to $\sin^2{(2 \theta_{new} )} \le 1.1\times 10^{-2}$ and
 $\sin^2{(2 \theta_{new} )} \le 2.2\times 10^{-2}$ respectively at
90\% and 99\% CL. Limits correspond to 
3.41 and to 7.13  events. }
\label{fig:limits1}
\end{figure}

\begin{figure*}[tbh]
\centering
\includegraphics[width=0.9\textwidth]{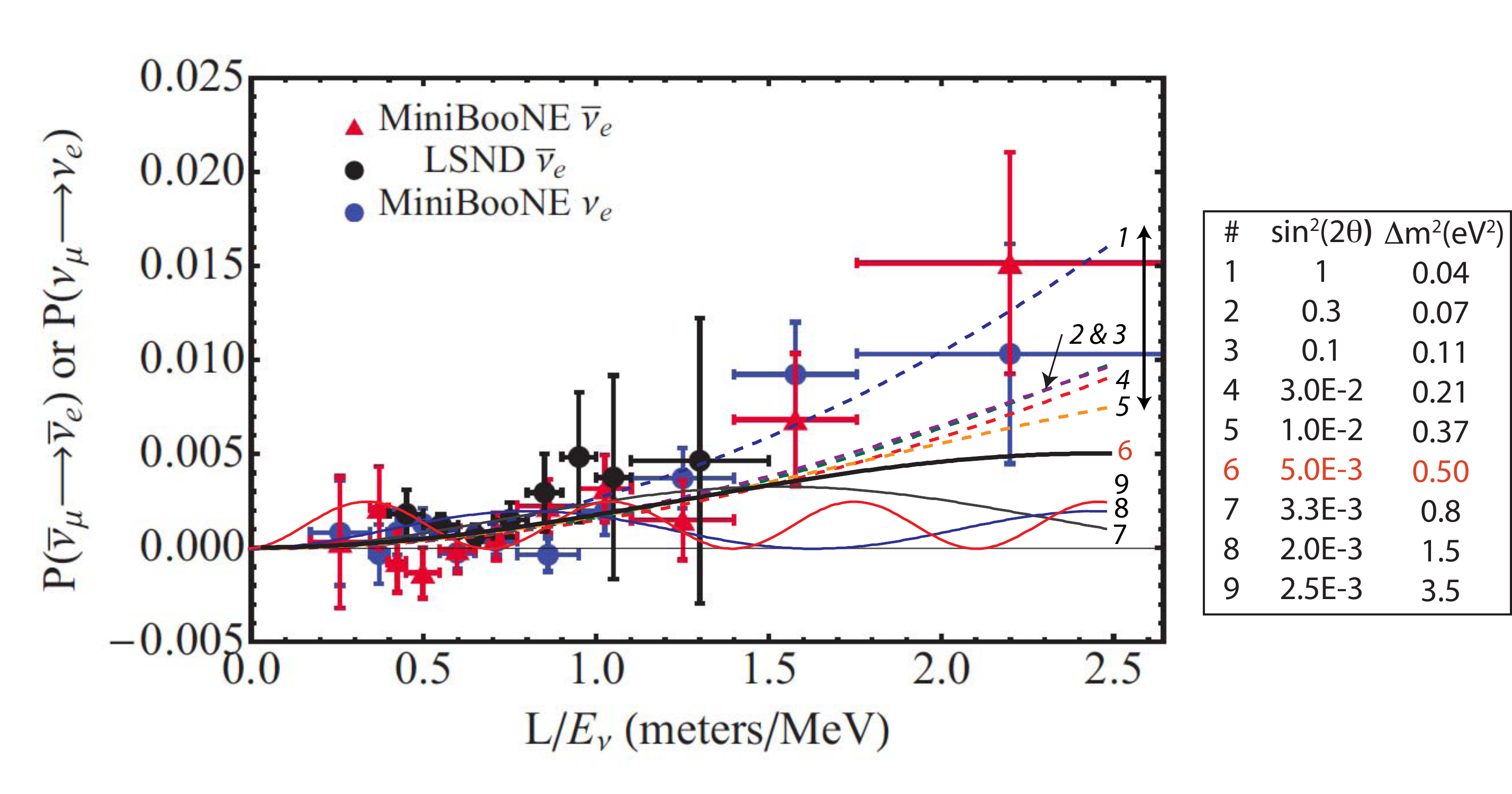}
\caption{Observed values of the LSND and MiniBooNE results are given
  with $\left< P_{\nu_\mu \rightarrow \nu_e}\right> $
 as a function of the distance $L/E_\nu$. 
The lines are examples of oscillation patterns with   sets of parameters
chosen within the MiniBoone allowed region \protect\cite{MiniBoone}. In particular, line~1
corresponds to the MiniBoone best fit in the combined  3+1 model\protect\cite{MiniBoone}.  
All lines are  consistent with  data at low
$L/E_\nu$ values. Solid lines, labeled from 6 to 9, are also
compatible  with the present ICARUS result. Instead,
parameter sets  indicated by 1-5 (dashed lines),  are driven  by
the  additional signal recorded by MiniBooNE for $L/E_\nu > 1$~m/MeV,
but they 
 are entirely ruled out by the present result because  they would
imply an excessive oscillation probability at the large  $L/E_\nu$  values
investigated by ICARUS.
Line 6 shows the ``best value'' including ICARUS results, with $ \left ( \Delta m^2, \sin^2{(2\theta )}\right )_{new} = (0.5\ 
\mathrm{eV}^2, 0.005) $.
 }
\label{fig:curves} 
\end{figure*}

\begin{figure*}[tbh]
\centering
\resizebox{0.95\textwidth}{!}{\includegraphics{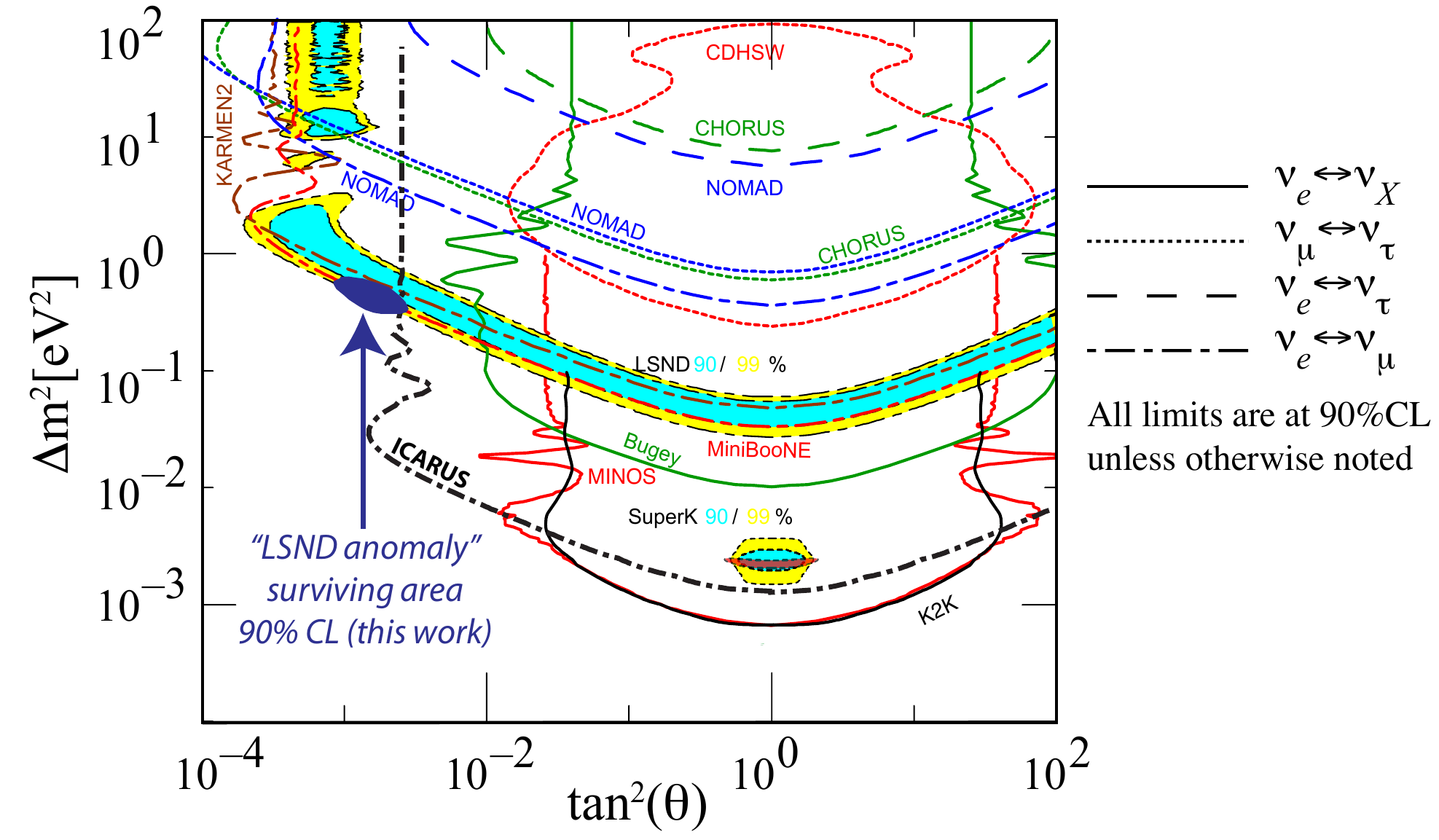}}
\caption{Regions in the $ \left ( \Delta m^2, \tan^2{(\theta )}\right )$ plane excluded 
by the ICARUS experiment compared with the published results~\protect\cite{PDG}.
While for $\Delta m^2_{new} \gg 1 eV^2$ there is 
already disagreement for $\nu_\mu \rightarrow \nu_e$  between the allowed
regions from the published experiments, for $\Delta m^2_{new} \leq 1 eV^2$ 
the ICARUS result now allows to define a much smaller, narrower allowed region 
centered around $\left( \Delta m^2, \sin^2{(2\theta )}\right)_{new} = (0.5 eV^2 , 0.005)$
in which there is a 90\% C.L. overall agreement.}
\label{fig:limits2} 
\end{figure*}

Within the range of our observations, our result is compatible with
the absence of a LSND anomaly.  Following Ref.~\cite{Stat}, at
statistical confidence levels of 90\% and 99\% and taking into account
the detection efficiency $\eta$, the limits due to the LSND anomaly
are respectively 3.4 and 7.1 events. According to the above described
experimental sample and the number of recorded events, the
corresponding limits on the oscillation probability are $\left<
P_{\nu_\mu \rightarrow  \nu_e}\right> = 5.4 \times 10^{-3}$
and $\left< P_{\nu_\mu \rightarrow  \nu_e}\right> = 1.1
\times 10^{-2}$ respectively.  The exclusion area of the ICARUS
experiment is shown in Figure~\ref{fig:limits1} in terms of the
two-dimensional plot of $\sin^2{\left( 2 \theta_{new}\right)}$ and
$\Delta m^2_{new}$.  In most of the area covered by ICARUS and allowed
by LSND and MiniBooNE, the oscillation averages approximately to a half
of its highest value, $\sin^2{\left(1.27\Delta m^2_{new} L/E_\nu
  \right )} \approx 1/2$. For lower values of $\Delta m^2_{new}$, the
longer baseline strongly enhances the oscillation probability with
respect to the one of the short baseline experiments. In ICARUS and
for instance with $\left(\Delta m^2 , \sin^2(2 \theta) \right)_{new}$ =
(0.11 eV$^2$, 0.10) as many as 30 anomalous $\nu_\mu \rightarrow \nu_e
$ events should have been present with $E_\nu \le 30$~GeV in the analysed sample.

The present result strongly limits the window of options from the
MiniBooNE experiment. Using a likelihood-ratio technique
\cite{MiniBoone}, CP conservation and the same oscillation
probability for neutrinos and antineutrinos, a best MiniBooNE
fit for Quasi Elastic (QE) events in the energy range  
$ 200$~MeV $< E_\nu ^{QE} <3000 $~MeV has been given at 
$ \left(  \Delta m^2 , \sin^2(2 \theta ) \right)_{new} $ = \hfill 
$\left( 0.037 \ \mathrm{eV}^2,1.00 \right)$.
 This is clearly excluded by the ICARUS result.  A 3+2
joint oscillation fit as a function of $E_\nu ^{QE}$ in both neutrino
and antineutrino modes has also been reported~\cite{MiniBoone} with
best fit values $\Delta m^2_{41}= 0.082 \ \mathrm{eV}^2$, $\Delta
m^2_{51}= 0.476 \  \mathrm{eV}^2$, 
$ \left | {\bf U}_{e,4}\right |^2  \left |{\bf U}_{\mu,4}
 \right |^2 = 0.1844$,
 $ \left |{\bf U}_{e,5} \right |^2 \left | {\bf U}_{\mu,5}\right |^2 =
 0.00547$. 
Also in this case the MiniBooNE value of $\Delta
m^2_{41}$ is clearly incompatible with the present ICARUS
result.

The oscillation probabilities from LSND are in the $L/E_\nu \le 1 $~m/MeV
region. The MiniBooNE result has extended the data to additional
values in the region $L/E_\nu \ge 1$~m/MeV (Figure~\ref{fig:curves}), corresponding to a significant
signal peak at smaller values of $E_\nu$.  
The actual origin of the excess may need further clarification, 
as already pointed out by the MiniBooNE Collaboration and for instance 
by Giunti and Laveder~\cite{Giunti}. In the low mass peak region the dominant 
signal is due to $\nu_\mu$ misidentified background adding to the observed LNSD signal.

As already mentioned,  the present experiment explores much larger values of $L/E_\nu$, 
but the ICARUS results
exclude also a substantial fraction of the MiniBooNE $ \left( \Delta
m^2 , \sin^2(2 \theta ) \right)_{new}$ curves shown in
Figure~\ref{fig:curves}, in particular the ones labeled  from 1 to 5.

A detailed comparison among  the various results on  
different oscillation phenomena, between different pairs of neutrino flavours, each having specific mixing angles and  $\Delta m^2$
is shown in Figure~\ref{fig:limits2}~\cite{PDG}. 
Even if disappearance and appearance results should not be  referred to a single effective $\theta$ and $\Delta m^2$, 
the plot allows situating the residual "LSND anomaly" in the framework of  the present neutrino oscillation results.  
  While for $\Delta
m^2_{new} >> 1$~eV$^2$ there is already disagreement between the allowed
regions from the published experiments, for $\Delta
m^2_{new} \le 1$~eV$^2$ the ICARUS result now
allows to define a much smaller, narrower region centered around $ \left( \Delta
m^2 , \sin^2(2 \theta ) \right)_{new}$=
(0.5 eV$^2$,  0.005) in which there is 90\% CL agreement between (1) the present 
ICARUS limit, (2) the limits of KARMEN and (3) the positive signals of LSND 
and MiniBooNE collaborations. This is the area in which the expectations from 
cosmology suggest a substantial contribution to the dark mass signal.

This region will be better explored by the proposed ICARUS/NESSiE dual
detector experiment~\cite{ICARUS-NESSIE} to be performed at
CERN at  much shorter distances ($\sim$300~m and $\sim$1.6~km) and lower neutrino
energies, which increase the event rate, reduce the overall
multiplicity of the events, enlarge the angular range and therefore
improve substantially the $\nu_e$ selection efficiency.

\section*{Acknowledgements}
The ICARUS Collaboration acknowledges the fundamental contribution to
the construction and operation of the experiment given by
INFN and, in particular, by the LNGS Laboratory and its Director.
The Polish groups
acknowledge the support of the Ministry of Science and Higher
Education , and of National Science Centre, Poland.
Finally, we thank CERN, in particular the CNGS staff,
for the successful operation of the neutrino beam facility.

%

\begin{thebibliography}{}
\bibitem{Pontecorvo}B. Pontecorvo, Zh. Eksp. Teor. Fiz. 53, 1717 (1967) [Sov. Phys. JETP 26, 984 (1968)].
\bibitem{LSND}A. Aguilar et al. 
Phys. Rev. D 64, 112007 (2001).
\bibitem{MiniBoone}A. A. Aguilar-Arevalo et al.,
  arXiv:1207.4809v1 [hep-ex] 19 Jul 2012 and references therein.
\bibitem{Reactors}	G. Mention et al.,  Phys.Rev. D83 (2011) 073006 and references therein.
\bibitem{MegaCurie} J. N. Abdurashitov et al., 
Phys. Rev. C 80, 015807 (2009).  
\bibitem{Gallex} F. Kaether, W. Hampel, G. Heusser, J. Kiko, and T. Kirsten, Phys. Lett. B 685, 47 (2010) and references therein.
\bibitem{ICARUS-INAUGURAL} C. Rubbia et al., JINST 6 P07011 (2011) and references therein.%
\bibitem{ICARUS-BIBBIA}  S. Amerio et  al., Nucl. Instr. and Meth. A527, 329 (2004).
\bibitem{CNGS} G. Aquistapace et al. CERN~98-02, INFN/AE-89-05 (1998); R. Bailey et al. CERN-SL/99-034 (DI), INFN/AE-99/05 Addendum (1999); E. Gschwendtner et al., CERN-ATS-2010-153 (2010).
\bibitem{CNGS-sim} A. Ferrari et al., CERN-AB-Note-2006-038 (2006).
\bibitem{FLUKA1} A. Ferrari et al., CERN-2005-10, INFN/TC-05/11,  (2005).
\bibitem{FLUKA2} G. Battistoni et al., AIP Conf. Proc. 896, 31-49,  (2007).
\bibitem{nufluxweb} http://www.mi.infn.it/$\sim$psala/Icarus/cngs.html
\bibitem{Atherton} H. W. Atherton et al., CERN Yellow Report 80-07 (1980). 
\bibitem{SPY}  G. Ambrosini et al., 
Eur. Phys. J. C10  605 (1999). 
\bibitem{Collazuol}	 G. Collazuol et al., Nucl. Instr. and Meth. A449,  609 (2000).
\bibitem{NA49} C. Alt et al., Eur. Phys. J. C 49, 897, (2007).	 
\bibitem{CNGS1} A. Ferrari et al, Nucl. Phys. B (Proc. Suppl.) 168, 169 (2007).
\bibitem{Charpak} G. Charpak et al., Nucl. Instrum. and Meth. 80, 13 (1970).
\bibitem{nutof1} M.~Antonello  et al., 
  Phys.\ Lett.\ B  713, 17 (2012).
\bibitem{3Dpaper} M. Antonello et al, 
 arXiv:1210.5089, accepted for publication in Advances in High Energy
 Physics (2013)
\bibitem{FLUKA-nu} G. Battistoni et al., Proceedings of the 12th International  conference on nuclear reaction mechanisms, Varenna (Italy), June 15-19,  2009, p.307.
\bibitem{50l} F. Arneodo et al., Phys. Rev. D74, 112001 (2006).
\bibitem{LAR1} S. Amoruso et al, 
Nucl. Instr. Meth. A523,  275 (2004).
\bibitem{LAR2} A. Antonello et al., 
Phys. Lett. B 711, 270 (2012).
\bibitem{FOGLI:t13}  G. L. Fogli et al. 	arXiv:1205.5254 [hep-ph] (2012);
\bibitem{PDG}J Beringer et al. (Particle Data Group), Phys. Rev. D86, 010001 (2012).
\bibitem{Stat} G.J. Feldman and R. D. Cousins, Phys. Rev. D 57 (1988) 3873.
\bibitem{KARMEN} B. Armbruster et al.,
Phys. Rev. D 65, 112001 (2002).
\bibitem{NOMAD} P. Astier et al.,
Phys. Lett. B 570  19 (2003).
\bibitem{CCFR} A. Romosan et al., 
Phys. Rev. Lett. 78 2912  (1997).
\bibitem{NuTeV} S. Avvakumov et al., 
Phys. Rev. Lett. 89 (2002) 011804.
\bibitem{Giunti} C. Giunti and M. Laveder, Phys. Rev. D 82,  053005 (2010).
\bibitem{ICARUS-NESSIE} M. Antonello et al., SPSC-P-347 (2012);
  C. Rubbia et al., SPSC-P-345 (2011). 
\end{thebibliography}
%
 \bibliographystyle{}
 
\end{document}